# Monte Carlo simulation of defects in hard-sphere crystal grown on a square pattern

**Atsushi Mori*, Yoshihisa Suzuki and Shigeki Matsuo**
*Institute of Technology and Science, University of Tokushima, 2-1 Minamijosanjima, Tokushima 770-8506, Japan*
**E-mail: mori@opt.tokushima-u.ac.jp*

**Abstract**

Monte Carlo simulations of hard-sphere (HS) crystal grown on a square patterned wall under gravity have been performed. While previous simulations were performed with step-wise controlled gravity, in the present simulations constant gravity has been applied from the first. In the case in which a flat wall is used as the bottom wall, if a large gravity is suddenly applied, the system does polycrystallize. On the other hand, in the present simulations, despite the sudden application of gravity, the system has not polycrystallize. Crystalline nucleation on the square pattern and successive crystal growth upward are suggested to overcome the homogeneous nucleation inside and result in. Defect disappearance, which has been essentially the same as that for the case with step-wise controlled gravity, has also observed for the present case. The characteristic of the square patterned bottom wall simulation with a large horizontal system size has been existence of triangular defects suggesting stacking tetra-hedra.

## 1. Introduction

The crystal/fluid phase transition in the hard-sphere (HS) system is sometimes referred to as Kirkood-Alder-Wainwright transition (or, shortly, Alder transition), which was discovered in 1957 by simulations (Wood and Jacobson, 1957; Alder and Wainwright, 1957). Because a phase transition in the system comprising of no attractive interaction is surprising, the Alder transition has so far attracted researchers. In 1968, Hoover and Ree determined the crystal/fluid coexistence region in the HS system as $\phi_f$ (= 0.494) < $\phi$ < $\phi_s$ (= 0.545) with $\phi = \left(\pi/6\right)\sigma^3\, N/V$ being the volume fraction of the HSs (Hoover and Ree, 1968). It should be noted that the phase diagram of the HS system does not depend on the temperature . In 1995, one of the present authors and coworkers successfully performed a direct simulation of the crystal/fluid coexistence (Mori *et al*., 1995). Davidchack and Laird further developed this simulation and revised the coexistence region as $\phi_f$ = 0.491 and $\phi_s$ = 0.542 (Davidchack and Laird, 1998).

In early years, the HS model was a mere idealized model for a colloidal dispersion. The inter-particle interaction between charge-stabilized colloid particles is well described by a repulsive Yukawa potential. Those systems were modeled by effective HSs; the phase transition into a colloidal crystal was explained as the Alder transition of effective HSs with the effective diameter being the core diameter plus an order of the Debye screening length (Wadachi and Toda, 1972). Nowadays, the HS model is not a mere idealized or effective mode; HS suspensions are synthesized such as poly (methyl methacrylate) (PMMA) spheres with surface polymer brushes dispersed in a hydrocarbon mixture (Antl *et al*., 1986). Such HS suspensions were extensively studied in view of the Alder transition and it was revealed that HS suspension exhibited the nature of the Alder transition (Pusey and van Megen, 1986; Paulin and Ackerson 1990; Underwood *et al*., 1994; Phan *et al*., 1996).

Interest in the photonic crystal, which arose in late 1980s, stimulated researches in the colloidal crystals. Theoretical predictions of photonic band were presented first by Ohtaka and then developed by Yablonobich and John (Ohtaka, 1979; Yablonovich 1987; John, 1987). Because a spatial periodic variation of dielectric constant with periodicity being of the same order of the light exists in the colloidal crystals, the colloidal crystals can exhibit a photonic band. To improve the photonic band structure a high crystallinity is required. Recently efforts were made to reduce the defect in the colloidal crystals. Sedimentation is one of the methods to produce the colloidal crystals; due to the densification by the sedimentation the colloidal dispersions does crystallize. Characteristic of the sedimentation method is that an equilibrium state in a sense is under manipulation. In the centrifugation sedimentation one can obtain a colloidal crystal more rapidly than in the gravitational sedimentation (2007; Suzuki *et al*., 2001).

An effect of gravity that reduces defects in colloidal crystals was found in 1997 (Zhu *et al*., 1997). The system was in a random hexagonal close pack (rhcp) structure under microgravity while the sediment was mixture of rhcp and face-centered cubic fcc crystals under normal gravity. This discovery was surprising because even if the stacking sequence varies the density does not change. We performed Monte Carlo (MC) simulations of HSs confined between well separated top and bottom flat walls under gravity (Mori *et al*., 2006). In those simulations a defective crystal was formed above a defectless crystal formed at the bottom for fcc (001) stacking. Transformation of the defective crystal into the defectless crystal was observed. By a close look at this process we found a glide mechanism for stacking fault shrinkage; an intrinsic stacking fault running along {111} oblique direction disappeared through the glide of a Shockley partial dislocation terminating the lower end of the stacking fault (Mori *et al*., 2007). In those simulations we varied the gravitational number $g^*\left(\equiv mg\sigma / k_B T\right)$ to avoid trapping of the system into a metastable state such as a polycrystalline state as reported by us previously (Yanagiya *et al*., 2005). Here, $m$ is the mass of a particle, $g$ is the acceleration due to gravity, $\sigma$ is the HS diameter, $k_B T$ is the temperature multiplied by the Boltzmann's constant. Those simulations had a shortcoming; the fcc (001) stacking was induced by the stress from a small lateral simulation box of a square shape. This shortcoming has been circumvented by use of a square patterned bottom wall (Mori, 2011bc). In those papers the driving force for the fcc (001) stacking has been replaced by the stress from the square pattern. The use of a template was first proposed by van Blaaderen *et al*. (van Blaaderen *et al*., 1997) and referred to as a colloidal epitaxy. In other words, the one of the present authors has succeeded in MC simulations of the colloidal epitaxy. In those simulations a square pattern, instead of fcc (001) pattern, has been used; this corresponds to Lin *et al*.'s experiment (Lin *et al*., 2000). The use of the square pattern has an advantage; by matching on the lattice line, instead of on the lattice point, the lattice constant of the crystal formed at the bottom is adjustable. So, in principle, the stress in the bottom crystal can be reduced.

Evolution of the center of the gravity for one of the previous simulations using a square pattern with the step-wise $g^*$ control is shown in Figure 1. Although overall equilibration is seen, the equilibration during an individual step at $g^*$ maintained at constant is not sufficient. An interest what results we would have if we relax the system keeping $g^*$ at a value for an enough time arose. We have at present performed MC simulations of the HS crystal on a square patterned substrate with a suddenly applied $g^*$.

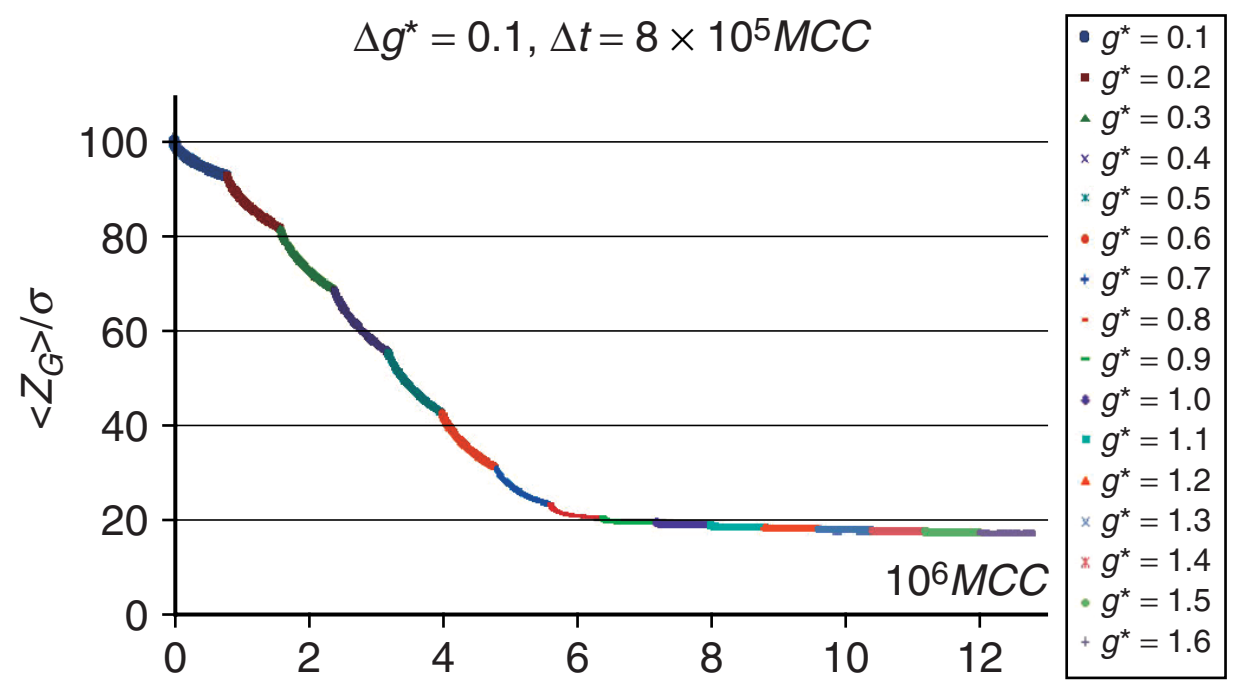

Fig. 1. Evolution of the center of the gravity for one of MC simulation of HSs confined between top flat and bottom square patterned walls with a step-wise control of $g^*$ $(\Delta g^* = 0.1, \Delta t = 8 \times 10^5 \mathrm{MCC}, N = 26624, L_x = L_y = 25.09\sigma, L_z = 200\sigma)$.

## 2. System and simulation

$N$ = 26624 HSs were confined in a simulation box of $L_x = L_y = 25.09\sigma$ with the periodic boundary condition and a pattered wall at $z = 0$ and a flat wall at $z = L_z = 1000\sigma$. The pattern on the bottom wall and the horizontal size of the bottom wall were the same as that for the previous one (Mori, 2011b). Grooves of width $0.707106781\sigma$ were formed in transverse and longitudinal directions with separation $0.338\sigma$ on the bottom wall. The separation between the neighboring groove centers was $1.045106781\sigma$. The diagonal length of the squares of the intersection of the grooves results in $0.9999999997\sigma$; the particle does not fall down to the bottom of the groove. For a schematic diagram of the pattern on the bottom, see a recent publication (Mori, 2011a).

Simulations at constant $g^*$ were continued for $5.12 \times 10^7$ Monte Carlo cycles (MCCs). Here, one MCC is defined such that it contains $N$ MC particle moves. The maximum displacement of particles was $\Delta r_{\max} = 0.6\sigma$, which was the same as previous simulations we performed (Mori, 2006, 2007, 2009, 2011bc). We employed the Metropolis's method with respect to the change of gravitational energy associated with the MC particle moves.

## 3. Results

Evolutions of the center of gravity for the simulations at constant $g^*$(=1.6, 1.1, 0.9, 0.6) are show in Figure 2. Moving block average over 1000 MCC are taken for the center of gravity $\langle z_G \rangle$. At the first stage the center of gravity sunk rapidly and then it reached at almost saturation. We show magnification of the almost flat part of Figure 2 in Figure 3. In any of Figure 3 a slight sink is seen. As shown in snapshots in the first stage formation of vacuum at the top of the system and simultaneous formation of crystalline phase at the bottom occurred. At an instant during the late stage defect disappearance accompanied with the sink of $\langle z_G \rangle$.

Let us look at snapshots at $g^* = 1.6$. Snapshots of a lower portion of the system at $g^* = 1.6$ at $1 \times 10^4$ and $1 \times 10^5$ MCCs are shown in Figure 4. As mentioned above the formation of vacuum at the top of the system and the crystallization at the bottom are seen. Snapshots at $1 \times 10^5$, $1.5 \times 10^5$, and $5 \times 10^7$ MCCs are shown in Figure 5. Comparing the top portion of Figures. 5(a) and (b) the crystallization is seen. This phenomenon is the same as that seen in Figure 4. Comparing the portions $5 < z/\sigma < 11$ of Figures 5(a) and (b) we find vanishing of the splitting of the projection of lattice lines, especially to the $xz$ plane. This corresponds to the disappearance of stacking fault; if a stacking fault runs crossing the lattice lines perpendicular to $xz$ plane, the projections of those lattice lines splits. This phenomenon has already observed in previous simulations with step-wise $g^*$ control (Mori, 2011bc). Also we already reported the disappearance of a staking fault in a flat wall case. We can say that stacking fault disappearance occurs

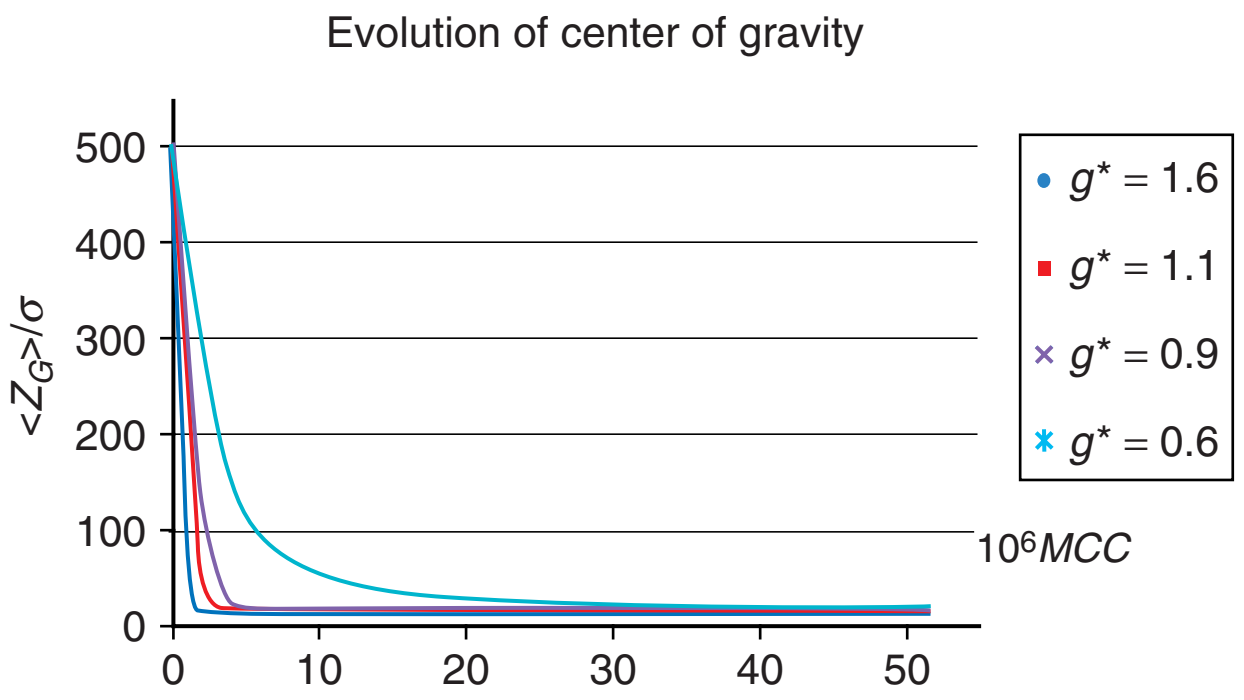

Fig. 2. Evolution of the center of the gravity for MC simulations of HSs confined between top flat bottom and bottom square patterned walls with a suddenly applied $g^*$.

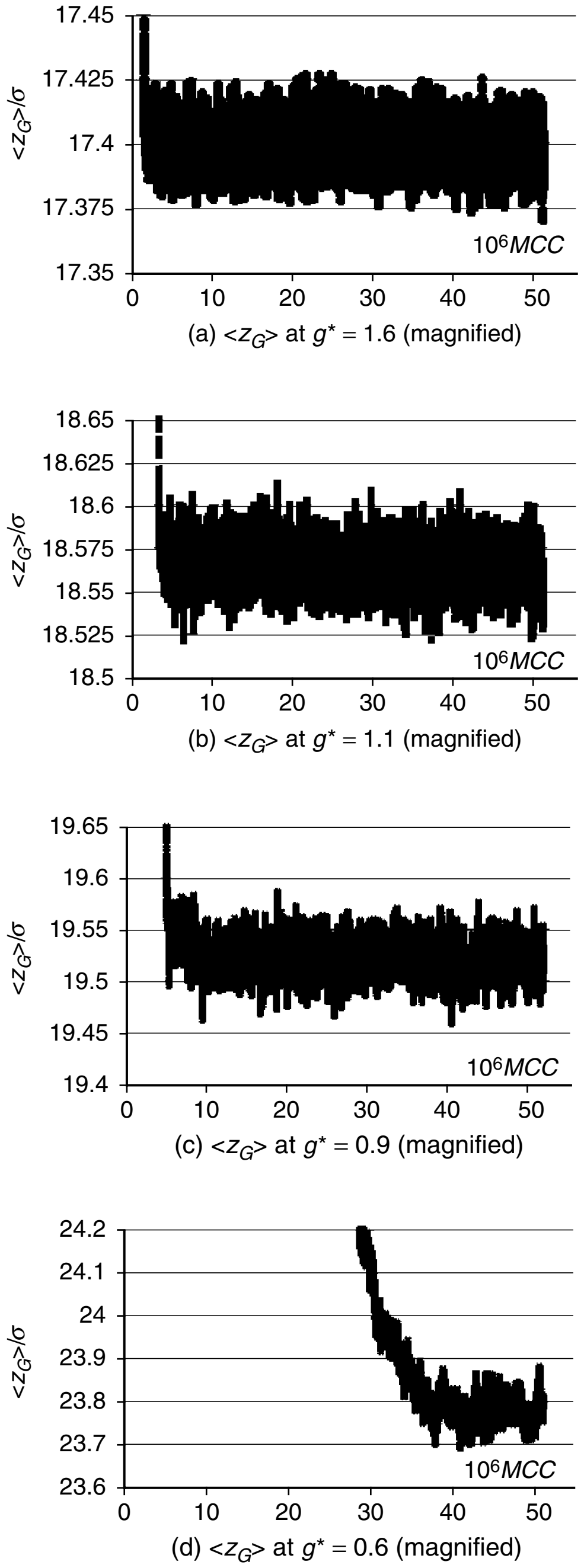

Fig. 3. Magnifications of the evolution of center of the gravity. The flat parts of Figure 2 are magnified.

both in HS crystal grown on a square pattern with step-wise controlled and suddenly applied $g^*$ and HS crystal grown in fcc (001) staking on a flat wall. We have already demonstrated the disappearance of stacking faults in fcc (001) stacking in a realistic condition; the present simulation is also the demonstration of the stacking fault disappearance in a realistic condition. The same phenomenon is seen in comparison between parts

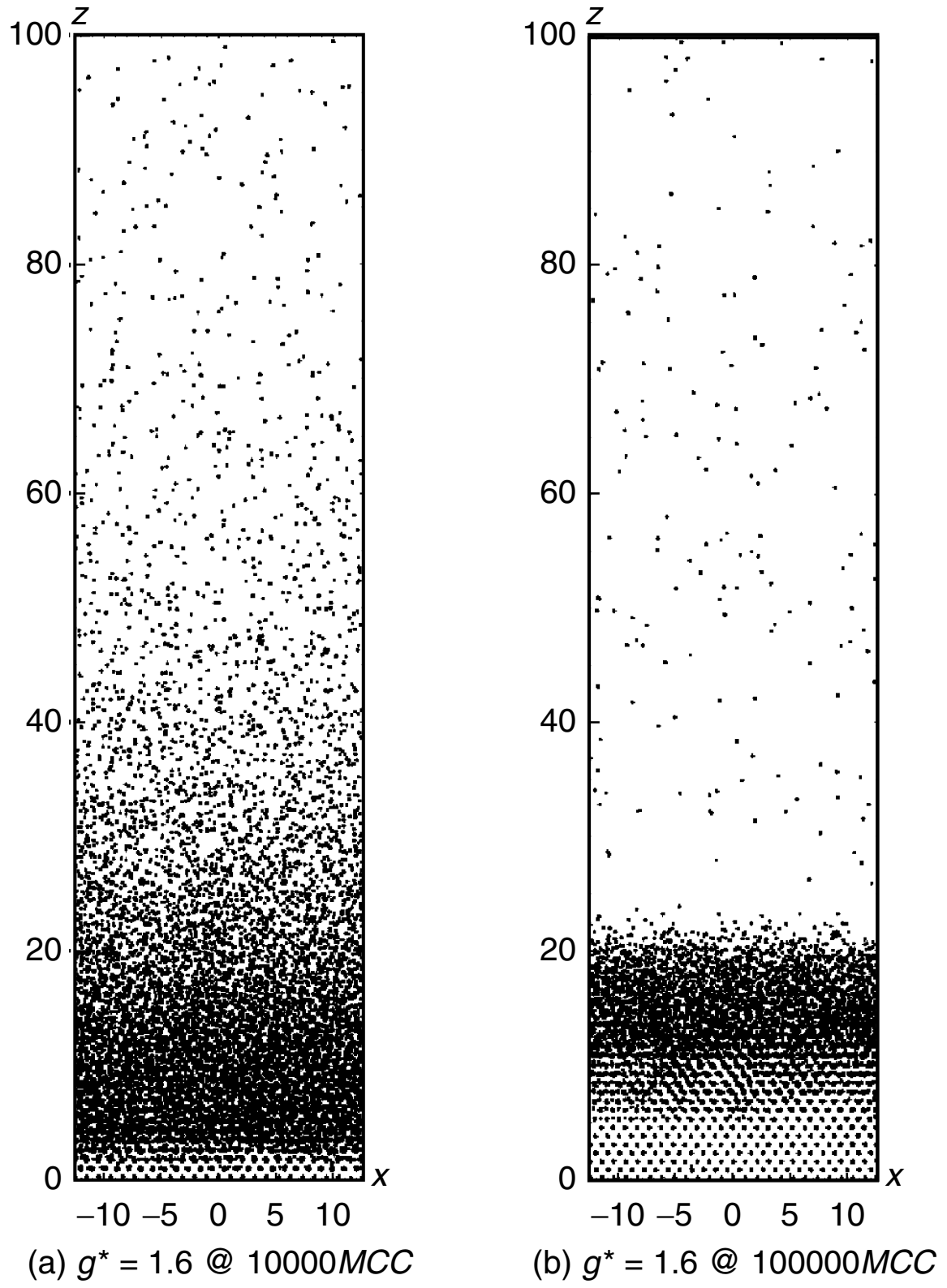

Fig. 4. Snapshots of lower part of the system at $g^* = 1.6$.

around $z/\sigma \sim 12.5$ and $y/\sigma \sim 1$ in *yz* projections in Figures. 5(b) and (c).

Another characteristic of Figures. 5(b) and (c) is existence of defects of a triangular shape in the projections. Though triangular structures have been seen in previous simulations with step-wise $g^*$ control, the triangular shapes have been more pronounced in the present simulations. While an upward triangle is seen around $z/\sigma \sim 8$ and $x/\sigma \sim 7$ in *xz* projection in Figure 5(c), downward triangles are seen around $z/\sigma \sim 8$ and $x/\sigma \sim 12$, $x/\sigma \sim 7$, and $x/\sigma \sim 5$ in *yz* projection in Figure 5(c). The upward and downward triangular shapes are understood by considering a tetrahedron surrounded by four fcc {111} face (in this consideration the faces which are converted by reversion with respect to the plane itself is regarded as identical). A projection of this tetrahedron onto one of {110} plane is an upward triangle and that onto the other {110} plane is a downward triangle. In this way, stacking fault tetrahedra are suggested.

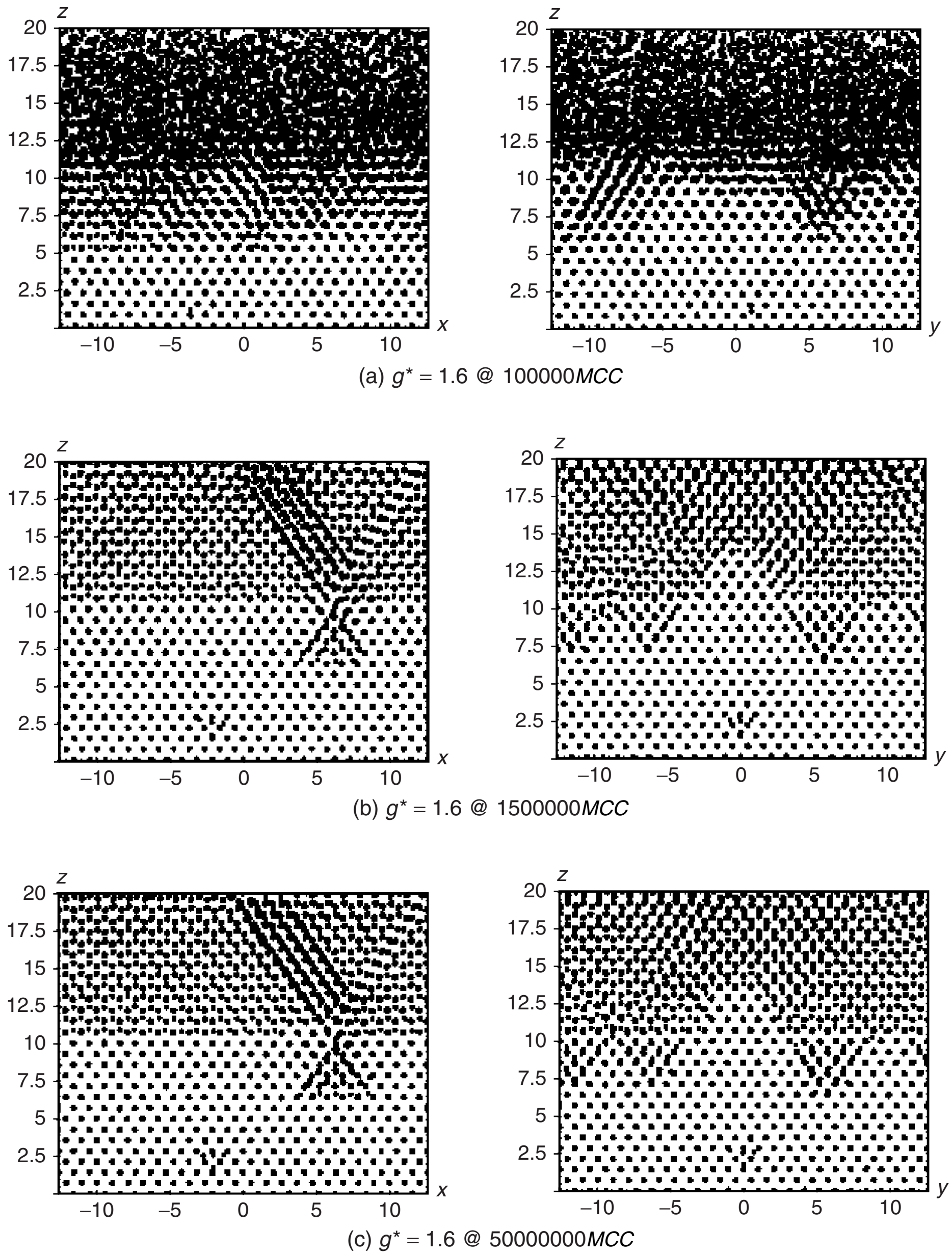

Fig. 5. Snapshots of bottom part of the system at $g^* = 1.6$.

What should be emphasized is avoidance of metastable states such as polycrystalline states. Even if $g^* = 1.6$ was suddenly applied, the system did not polycrystallize.

The result for $g^* = 1.1$, 0.9 and 0.6 are essentially the same.

## 4. Discussions

One of the most notable phenomena is the avoidance of the polycrystallization. A crystal nucleation occurred on the patterned substrate and the upward crystal growth occurred in succession. If this growth overcame the homogeneous nucleation inside, then the polycrystallization could be avoided. Unlike the flat wall case the step-wise control of $g^*$ is unnecessary in the colloidal epitaxy. This indicates a superiority of the colloidal epitaxy.

We have, however, found stacking tetrahedra, which seem sessile. Why was this defect structure not observed in previous simulations of (001) stacking due to the stress from small simulation box (Mori, 2006; Mori, 2007)? It may be due to the smallness of the system. Indeed, we have seen triangular like shapes in the square pattern simulation with a small system size (Mori, 2011c), but conjecture of the existence of the stacking fault tetrahedra has been postponed. The stacking fault tetrahedra have been suggested in large system size simulations (Mori, 2011b).

## 5. Conclusion

In conclusion, we would like to say that we have successfully performed Monte Carlo simulations of fcc (001) growth of hard-sphere crystals in a realistic condition, too. In other words, we could perform Monte Carlo simulations of the colloidal epitaxy. Moreover, it is followed that the colloidal epitaxy has more two superiorities other than the original idea of the colloidal epitaxy. The original idea is the uniqueness in the stacking sequence in fcc (001) stacking. One of superiorities is the glide mechanism of the Shockley partial dislocation for disappearance of a stacking fault, already found in a previous work (Mori *et al.*, 2007). The other is avoidance of the polycrystallization.

Unlike the flat wall case (Yanagiya *et al.*, 2005), even if we suddenly apply the gravity such as $g^* = 1.6$ the system does not polycrystallize in the case of the squared patterned wall. This may be because the crystalline nucleation on the patterned substrate overcomes. Band structures of the stacking disorder, which were reported previously (Mori, 2011b), have been avoided by enlarging the lateral system size. Instead, stacking fault tetrahedra have been suggested (Mori, 2011a). Disappearance of stacking disorder is conjectured.

Stacking fault tetrahedra have been suggested. So, at a glance the colloidal epitaxy has a fault. As discussed above, however, it may be due to the smallness of the system size that in small system simulations no stacking fault tetrahedra were formed. In a large system stacking disorders running along various directions may collide with each other or go across. As a result, complex defect structures form in a large system.

In the present study $g^*$ has been maintained at constant. In previous work with the step-wise controlled $g^*$ (Mori, 2011b), the stacking fault tetrahedra were not disappeared. Nevertheless, we expect more sophisticated $g^*$ control enables the stacking fault tetrahedra disappearance.